\documentclass[superscriptaddress,nofootinbib,prb,amsmath,twocolumn]{revtex4}
\usepackage{times}
\usepackage{hyperref}
\usepackage{graphicx}
\usepackage{longtable}
\usepackage{color}
\usepackage{amssymb}

\newcommand{\beq}{\begin{equation}}
\newcommand{\eeq}{\end{equation}}
\newcommand{\ket} [1] {\vert#1\rangle}

\def\opone{\leavevmode\hbox{\small1\kern-3.8pt\normalsize1}}

\begin{document}

%%%%%%%%%%%%%%%%%% title page information %%%%%%%%%%%%%%%%%%
\title{Polarization entanglement with GRaded-INdex lenses}

% \email{opex@osa.org} %% email affiliation is required


\author{Giuseppe Vallone}
\email{giuseppe.vallone@uniroma1.it} %% email affiliation is required
\homepage{http://quantumoptics.phys.uniroma1.it/}
\affiliation{Centro Studi e Ricerche ``Enrico Fermi'', Via Panisperna 89/A, Compendio del Viminale, Roma 00184, Italy}
\affiliation{Dipartimento di Fisica, ``Sapienza''
Universit\`{a} di Roma,
Roma 00185, Italy }
\author{Gaia Donati}
\homepage{http://quantumoptics.phys.uniroma1.it/}
\affiliation{Dipartimento di Fisica, ``Sapienza''
Universit\`{a} di Roma,
Roma 00185, Italy }
\author{Francesco De Martini}
\homepage{http://quantumoptics.phys.uniroma1.it/}
\affiliation{Dipartimento di Fisica, ``Sapienza''
Universit\`{a} di Roma,
Roma 00185, Italy }
\affiliation{Accademia Nazionale dei
Lincei, via della Lungara 10, Roma 00165, Italy}
\author{Paolo Mataloni}
\homepage{http://quantumoptics.phys.uniroma1.it/}
\affiliation{Dipartimento di Fisica, ``Sapienza''
Universit\`{a} di Roma,
Roma 00185, Italy }
\affiliation{Istituto Nazionale di Ottica Applicata (INOA-CNR),
L.go E. Fermi 6, 50125 Florence, Italy}

% \pacs{..}

\begin{abstract}
By using an optical device based on the integration of a GRaded-INdex (GRIN) rod lens and a single-mode optical fiber we efficiently
collected photon pairs generated by spontaneous parametric down-conversion.
We show that this system preserves the polarization entanglement of the 2-photon states.
Hence this device, characterized by a remarkable easiness of alignment and allowing for high coupling 
efficiency of single-mode radiation, can be used with photons entangled in various degrees of freedom, 
such as polarization and spatial momentum,
and to interconnect different sides of complex optical circuits.
\end{abstract}

\maketitle

%%%%%%%%%%%%%%%%%%%%%%%%%%  body  %%%%%%%%%%%%%%%%%%%%%%%%%%
% \section{Introduction}
Photon entanglement is a key resource for
modern quantum information (QI) applications such as quantum
computation \cite{knil01nat,raus01prl} and communication protocols \cite{benn84ieee, eker91prl}. 
Spontaneous parametric down-conversion (SPDC) is up to now the
most established technique to generate photon entanglement by using polarization and other 
degrees of freedom (DOFs) of the 
photons\cite{kwia95prl1, kwia99pra, rari90prl,mair01nat, ross09prl, barr05prl,naga09prl, fran89prl,bren99prl}.
% such as polarization \cite{kwia95prl1, kwia99pra},
% linear momentum (either transverse or longitudinal) \cite{rari90prl,mair01nat,ross09prl}, 
% orbital angular momentum \cite{mair01nat,barr05prl,naga09prl} 
% and time-energy (or time-bin)\cite{fran89prl,bren99prl}. 
The realization of four\cite{walt04nat} and six-photon\cite{lu07nap} entangled states
together with the introduction of hyperentangled states \cite{kwia97jmo,barb05pra, vall07prl}, based on the
entanglement of two photons in many DOFs (such as
polarization and longitudinal momentum),
open important perspectives in the implementation of QI tasks with multiqubit photon states.

In a recent paper \cite{ross09prl}, we have demonstrated the entanglement of two
photons in many optical modes by collecting the
degenerate conical SPDC radiation (with horizontal polarization) of a Type I phase-matched
crystal. Multi-path entanglement was realized by a set of graded-index (GRIN)
lenses, carefully aligned in the annular section of the SPDC cone
and efficiently connected to a bundle of single-mode optical
fibers. This experiment proved the feasibility of using GRIN lenses with longitudinal momentum entanglement
and thus suggests to test the entanglement in other DOFs, such as polarization. 
In particular the absence of any birefringence effect, 
introducing a polarization decoherence due to temporal and/or spatial walk-off in the GRIN lenses, should be verified.
\begin{figure}[htbp]
\centering\includegraphics[width=8cm]{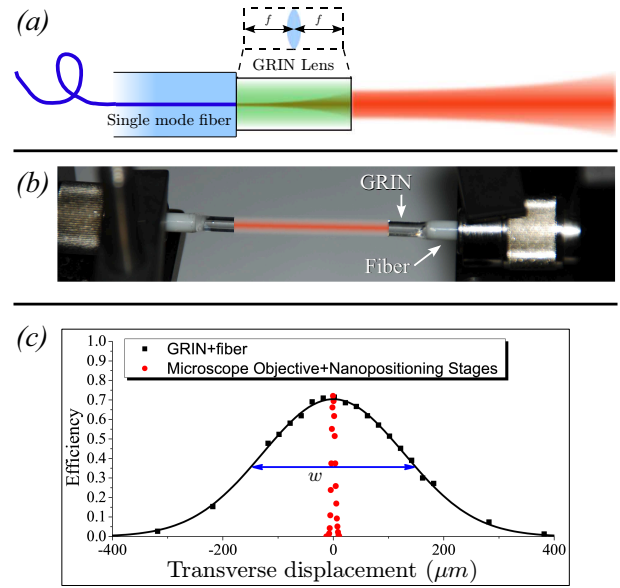}
\caption{$(a)$ Sketch of the GL-SMF system lying within the confocal parameter of the input mode. 
 $(b)$ Picture of two GL-SMF systems put face to face. 
 Each GRIN lens is glued to the input face of the corresponding
 single-mode optical fiber.
(c) Comparison between the collection efficiency of a GL-SMF (black curve) and a 40x objective as a function of  transverse displacement.
$w=(301\pm4)\mu m$ is the full width at half maximum of the gaussian fitting curve.
} \label{fig:GRIN}
\end{figure}

In this paper, we investigate the polarization performances of the integrated device composed 
of a GRIN rod lens and a single-mode optical fiber.
We focus on the 
% main features of this optical device as well as testing and successfully proving the 
preservation of polarization entanglement. 
This result, together with the high collection efficiency, the easiness 
of alignment and the low translational sensitivity, makes 
this optical integrated system a powerful tool for advanced QI tasks.

% \section{Experimental setup}
% \subsection{Preparation of the integrated optical system} 
The basic elements of the optical device we are going to consider are
a single-mode fiber (SMF; Thorlabs\cite{thorlabs}, mod. P1-630A-FC-2) and a ``quarter-pitch'' GRIN rod lens (GL;
Grintech\cite{grintech}, mod. GT-LFRL-200-025-50-NC(728), length = 5.0 mm,
diameter = 2.0 mm, numerical aperture = 0.5, AR-coated on the
input face). 
We recall that the term ``quarter-pitch'' refers to the GRIN lens length, chosen in such a way that an input plane 
wave is focused on the output face of the GRIN lens.
\begin{figure*}[t]
\centering\includegraphics[width=15cm]{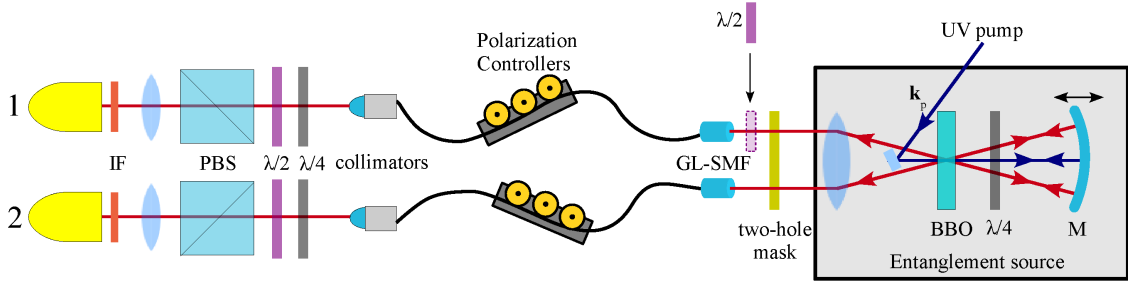}
\caption{Experimental setup used for polarization Bell-state measurements.
The basic elements of the entanglement source are described in Ref. \cite{cine04pra}} \label{fig:setup}
\end{figure*}

The GL is glued to one end of the fiber to give the
definitive integrated optical system, from here on labeled as
GL-SMF. The coupling efficiency of a GL-SMF system is essentially determined during the 
fixing procedure of the two components. The latter was performed by using a continuous wave (cw), 
horizontally-polarized Ti:Sapphire laser beam ($\lambda_{Ti:Sa} = $728 $nm$) which simulates
the input radiation collected by the two GL-SMF systems included in the experimental
setup. A positive lens
(focal length 500 $mm$) was inserted in the Ti:Sa
beam path to model the beam shape and let the beam waist coincide
with the location of the GL.
This choice approximately sets the optimal coupling conditions for the radiation entering the GL-SMF system.
As shown in Fig. \ref{fig:GRIN}(a), the considered GRIN lens is equivalent to a lens with focal length 
$f\simeq2mm$\footnote{Here $f=\frac{1}{n_0g}$, where $n_0$ is the maximum value of the index of refraction 
within the GRIN lens and $g=0.312mm^{-1}$ is its gradient constant\cite{grintech}.}
and two free space propagations of length $f$. Hence it is possible to describe
the coupled beam in terms of its beam waist $W'_0\simeq\frac{\lambda}{\pi W_0}f\simeq185 \, \mu m$ 
and the corresponding confocal parameter $z'_0=\frac{\pi W'^2_0}{\lambda}\simeq150 \, mm$.
These values can be calculated from the core diameter of the fiber ($2W_0\simeq5 \, \mu m$, see \cite{thorlabs}).

First, we aligned the input face of the GL orthogonal to the laser beam, 
with its AR-coated side turned to the input radiation in order to avoid reflections.
We then brought the SMF closer to the GL and optimized the coupling into the fiber.
By knowing the input power, it was
possible to obtain the optimal coupling conditions by monitoring
the collection efficiency of each GL-SMF system (calculated as
$P_{out}$/$P_{in}$) and improve it by sub-micrometric
displacements and careful tilting of the fiber. Once the GL and
the SMF were sufficiently close to each other, we proceeded to glue
the two components. A coupling efficiency of almost 70\% and a gaussian shape translational efficiency with
full width at half maximum equal to $w=(301\pm4)\mu m$ were
measured for the two integrated systems built by this procedure.

% \subsection{Description of the source of polarization-entangled states} 
Successively, the two equal GL-SFM systems were used to collect polarization-entangled photon pairs
generated by a SPDC source.
The source, shown in Fig. \ref{fig:setup}, and described in another paper\cite{cine04pra},
generates the polarization-entangled state 
\beq \ket{\Phi^\theta} =
\frac{1}{\sqrt{2}}({\ket{H}}_1{\ket{H}}_2 +
e^{i\theta}{\ket{V}}_1{\ket{V}}_2)\,, 
\eeq 
over two modes selected from the degenerate emission cone of a Type I $\beta$-Barium-Borate (BBO) crystal.
The selected SPDC radiation was then collected by the two GL-SMF
systems. For each SMF, a manual
polarization controller (PC) allowed to
invert the unitary transformation introduced by the fiber.
Finally, two standard polarization-analyzer settings constituted by a
$\lambda/4$ wave-plate, a $\lambda/2$ wave-plate and a polarizing
beam splitter (PBS) allowed the characterization of the output
radiation focused on two single-photon detectors (Perkin-Elmer,
mod. SPCM-AQR14), labeled as 1 and 2 in Fig. \ref{fig:setup}. Two interference filters (IF; bandwidth
$\Delta\lambda$) placed in front of the detectors determine the
coherence time of the emitted photons; in the present experiment
we considered two different values of the bandwidth,
namely $\Delta\lambda = 6 \, nm$ and $\Delta\lambda = 70 \, nm$. With the former (latter) choice 
we obtained nearly $180 \, coinc/s$ (1000 $coinc/s$) within a coincidence window of $3 \,ns$.
This determines a coincidences/singles ratio of almost $8\%$.

\begin{figure}[h]
\begin{center}
\includegraphics[width=8cm]{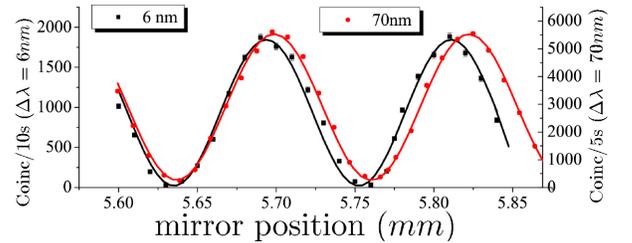}
\caption{Coincidence rate measured as a function of the mirror M
position for $\Delta\lambda = 6 \, nm$ (black curve) and $\Delta\lambda = 70 \, nm$ (red curve).} \label{fig:osc}
\end{center}
\end{figure}

% \section{Experimental results} 
By varying the position of the spherical mirror M (shown in Fig. \ref{fig:setup}), we were able to change
the relative phase $\theta$ in the generated state
$\ket{\Phi^\theta}$. We measured the photon coincidence rate as a
function of the mirror position with the polarization analyzers
set to diagonal polarizations, i.e. we detected both photons
with polarization $\frac{1}{\sqrt2}(\ket{H}+\ket{V})$. The
experimental results are shown in Fig. \ref{fig:osc} for the two bandwidth values 
$\Delta\lambda$ of the IFs. In our experimental conditions, 
the $\Delta\lambda = 70 \, nm$ IFs were used to filter out stray-light radiation. 
In this case, the actual bandwidth of the detected photons is narrower than 70 $nm$ because of two effects:
the spatial mode selection operated by the SMFs and the presence of position-frequency correlations characteristic
of the SPDC radiation.

%%%%%%%%%%%%%%%%%%%%%%%%%%%%%%%%%%%%%%%%%%%%%%%%%%%%%%%%%%%%%%%%%
% Figure
%%%%%%%%%%%%%%%%%%%%%%%%%%%%%%%%%%%%%%%%%%%%%%%%%%%%%%%%%%%%%%%%%
%%%%%%%%%%%%%%%%%%%%%%%%%%%%%%%%%%%%%%%%%%%%%%%%%%%%%%%%%%%%%%%%%
% %%%%%%%%%%%%%%%%%%%%%%%%%%%%%%%%%%%%%%%%%%%%%%%%%%%%%%%%%%%%%%%%%
% % Figure
% %%%%%%%%%%%%%%%%%%%%%%%%%%%%%%%%%%%%%%%%%%%%%%%%%%%%%%%%%%%%%%%%%
% \begin{figure}[h]
% \begin{center}
% \includegraphics[width=8cm]{osc_mirrorfixed}
% \caption{Measurement of polarization entanglement. The selected state is $\ket{\Phi^{+}}$.} \label{fig:osc_mirrorfixed}
% \end{center}
% \end{figure}
% %%%%%%%%%%%%%%%%%%%%%%%%%%%%%%%%%%%%%%%%%%%%%%%%%%%%%%%%%%%%%%%%%

The variation of the number of coincidences $N_C$ with the phase $\theta$
is precisely the one expected for entangled states such as
$\ket{\Phi^\theta}$,
$N_{C}(\ket{\Phi^\theta})\propto1+\cos\theta$. The coincidence
fringe visibility reaches the value $V=0.9785\pm0.0001$ for
$\Delta\lambda=6nm$, while $V=0.9094\pm0.0001$ for
$\Delta\lambda=70nm$. The maxima and minima of these interference
patterns single out the longitudinal positions of the mirror M
corresponding to the Bell states $\ket{\Phi^{+}} =
\frac{1}{\sqrt{2}}({\ket{H}}_1{\ket{H}}_2
+{\ket{V}}_1{\ket{V}}_2)$ and $\ket{\Phi^{-}} =
\frac{1}{\sqrt{2}}({\ket{H}}_1{\ket{H}}_2
-{\ket{V}}_1{\ket{V}}_2)$, respectively. 
The slight difference between the two fringe pattern periods in Fig. \ref{fig:osc}
is mainly due to the different bandwidth values. 
% By selecting a specific Bell state we give a further proof of the presence of polarization entanglement. 
% Fig. \ref{fig:osc_mirrorfixed} then shows the coincidence oscillations measured 
% for $\ket{\Phi^{+}}$ as a function of the orientation $\alpha$ of analyser $1$ with  analyser $2$ selecting 
% the polarization state $\frac{1}{\sqrt2}(\ket{H}+\ket{V})$. In this case, we obtain
% a visibility $V=0.993\pm0.002$ for $\Delta \lambda=6nm$ and $V=0.910\pm0.002$ for $\Delta \lambda=6nm$.

\begin{figure}[t]
\begin{center}
\includegraphics[width=7cm]{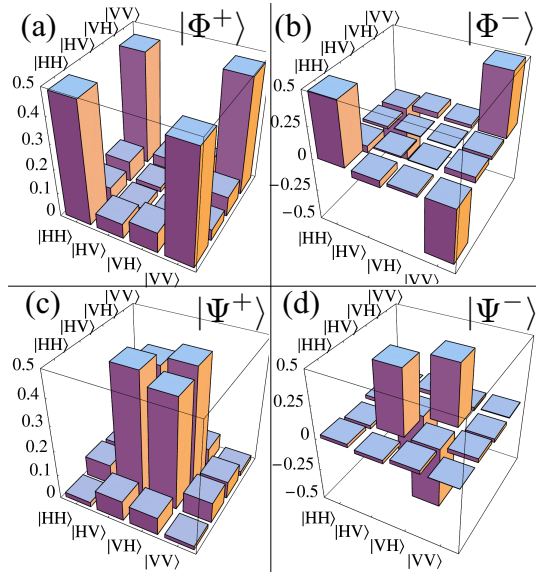}
\caption{\footnotesize{Tomographic reconstruction of the four
polarization-entangled generated states (real parts). A maximum
likelihood technique was used to obtain physical density matrices.
The imaginary components are negligible. The corresponding
theoretical Bell states are: (a) $\ket{\Phi^{+}}$, (b)
$\ket{\Phi^{-}}$, (c) $\ket{\Psi^{+}}$, (d) $\ket{\Psi^{-}}$.}}
\label{fig:tomo}
\end{center}
\end{figure}

The tomographic reconstruction of the coupled radiation represents the primary instrument to evaluate
the polarization performances of the GL-SMF setup. Besides the two
Bell states cited above, which are derived from the original
generated entangled state $\ket{\Phi^\theta}$ by appropriately
setting the mirror M position, we can obtain $\ket{\Psi^{\pm}}=
\frac{1}{\sqrt{2}}({\ket{H}}_1{\ket{V}}_2
\pm{\ket{V}}_1{\ket{H}}_2)
$ thanks to an additional $\lambda/2$ wave-plate placed in front of
one of the integrated GL-SMF systems. Fig. \ref{fig:tomo} shows the
results of the four tomographic characterizations, with the
elements of the density matrices written in the polarization basis
$\{\ket{HH},\ket{HV},\ket{VH},\ket{VV}\}$.

%%%%%%%%%%%%%%%%%%%%%%%%%%%%%%%%%%%%%%%%%%%%%%%%%%%%%%%%%%%%%%%%%
% Figure
%%%%%%%%%%%%%%%%%%%%%%%%%%%%%%%%%%%%%%%%%%%%%%%%%%%%%%%%%%%%%%%%%
%%%%%%%%%%%%%%%%%%%%%%%%%%%%%%%%%%%%%%%%%%%%%%%%%%%%%%%%%%%%%%%%%

For every generated state, we note the presence of the
characteristic diagonal terms and
the strong coherences (with negative signs when looking to
the singlet state $\ket{\Psi^{-}}$ and to $\ket{\Phi^{-}}$) existing between them. The fidelities of the
experimental states are: $F_{\ket{\Phi^{+}}} = 0.938 \pm 0.019$,
$F_{\ket{\Phi^{-}}} = 0.949 \pm 0.020$, $F_{\ket{\Psi^{+}}} =
0.923 \pm 0.014$ and $F_{\ket{\Psi^{-}}} = 0.965 \pm 0.029$, all
of which show a strong closeness to the theoretical Bell states
(encoded in polarization). A more quantitative parameter
associated to the generated polarization-entangled states is given
by the tangle $\tau$: $\tau_{\ket{\Phi^{+}}} = 0.873$,
$\tau_{\ket{\Phi^{-}}} = 0.940$, $\tau_{\ket{\Psi^{+}}} = 0.846$
and $\tau_{\ket{\Psi^{-}}} = 0.911$. These values are the evidence
of a high degree of entanglement generation.

% \section{Conclusions} 
We presented a device built as 
an integrated system of a graded-index lens glued to 
a single-mode optical fiber. In the present paper we have shown that this system
preserves polarization entanglement of SPDC radiation, 
despite possible anisotropies between the radial and tangential components of the mode\cite{tent03apo}.
Since this very device gave successful results when used with longitudinal 
momentum entanglement\cite{ross09prl}, it can be adopted for a series of QI applications 
where different types of entanglement are involved. Particularly, 
it could be a useful instrument to connect different sides 
of complex optical circuits.

When thinking of future 
applications involving optical microchips, it is worth investigating the possible
solutions to link distinct elements in order to perform more complex operations\cite{poli08sci, mars09ope, smit09qph}. 
For this purpose we studied the coupling efficiency of the sequence of two GL-SMF 
systems separated by a variable path in free space and set out as pictured in Fig. \ref{fig:GRIN}(b). 
For a free space separation 
changing from $1 \, mm$  to $40 \, mm$ we measured a coupling efficiency of almost $90\%$ . 

This work was supported by Finanziamento Ateneo 2008 (prot. C26A08RMYF) of 
Sapienza Universit\`a di Roma.

Thanks to L. Businaro and R. Ursin for useful discussions and to M. Figliuzzi and A. Rossi 
for their support in the preparation of the experiment.

 %   \bibliographystyle{../../../bibliografia/customize_bibtex/template/prl_with_titles}% Produces the bibliography via BibTeX.
%   \bibliographystyle{../../../bibliografia/customize_bibtex/template/prl_with_et_al}% Produces the bibliography via BibTeX.
%   \bibliography{../../../bibliografia/qi-bibliografia}% Produces the bibliography via BibTeX.

\end{document}